\documentclass[prb,twocolumn,showpacs,preprintnumbers,amsmath,amssymb,superscriptaddress,10pt]{revtex4-1}

\usepackage{graphicx}
\usepackage{dcolumn}
\usepackage{bm}

\def\Vec#1{\bm{#1}}

\begin{document}


\title{ Superconducting Gap Function in an Organic Superconductor (TMTSF)$_{2}$ClO$_{4}$ with Anion Ordering; First-principles Calculations 
and Quasi-classical Analyses for Angle-resolved Heat Capacity }

\author{Yuki Nagai}
\affiliation{CCSE, Japan  Atomic Energy Agency, 6-9-3 Higashi-Ueno, Tokyo 110-0015, Japan}
\affiliation{CREST(JST), 4-1-8 Honcho, Kawaguchi, Saitama, 332-0012, Japan}
\author{Hiroki  Nakamura}
\affiliation{CCSE, Japan  Atomic Energy Agency, 6-9-3 Higashi-Ueno, Tokyo 110-0015, Japan}
\affiliation{CREST(JST), 4-1-8 Honcho, Kawaguchi, Saitama, 332-0012, Japan}
\author{Masahiko Machida}
\affiliation{CCSE, Japan  Atomic Energy Agency, 6-9-3 Higashi-Ueno, Tokyo 110-0015, Japan}
\affiliation{CREST(JST), 4-1-8 Honcho, Kawaguchi, Saitama, 332-0012, Japan}


\date{\today}

\begin{abstract}

We calculate angle-dependent heat capacity in a low magnetic field range on 
the basis of Kramer-Pesch approximation together with an electronic structure obtained by 
first-principles calculations to determine a superconducting gap function 
of (TMTSF)$_{2}$ClO$_{4}$ through its comparisons with experiments. 
The present comparative studies reveal that a nodal $d$-wave gap function consistently explains 
the experimental results for (TMTSF)$_{2}$ClO$_{4}$. 
Especially, it is emphasized that 
the observed unusual axis-asymmetry of the angle-dependence eliminates 
the possibility of $s$-wave and node-less $d$-wave functions. 
It is also found that the directional ordering of ClO$_{4}$ anions does not have 
any significant effects on the Fermi surface structure 
contrary to the previous modelings since the 
two Fermi surfaces obtained by the band calculations almost cross within the present 
full accuracy in first-principles calculations. 

\end{abstract}

\maketitle
\section{Introduction}

The organic superconductors, (TMTSF)$_{2}$$X$'s have attracted 
considerable attention because of their rich variety of phases including superconducting states 
in spite of rather simple quasi-one-dimensional (Q1D) electronic structures. 
For example, (TMTSF)$_{2}$PF$_{6}$ exhibits a pressure induced superconductivity 
whose pairing symmetry has been expected to be spin-triplet by 
Knight-shift measurements \cite{Lee}. Meanwhile, 
(TMTSF)$_{2}$ClO$_{4}$ is an ambient pressure superconductor, whose 
Knight-shift measurements suggest a conventional spin-singlet pairing \cite{Shinagawa}. 
Since the singlet superconducting phase is in close proximity 
to spin density wave (SDW) states, spin-fluctuation has been proposed as a candidate of 
the pairing glue. 
However, the pairing mechanism is still far from an established settlement
for (TMTSF)$_{2}$$X$ in spite of several 
theoretical studies \cite{Kuroki, Aizawa, Tanaka, Rozhkov, Belmechri,Shimahara,KurokiAoki}.
 
Synchronous to the theoretical struggle, a number of experimental 
studies have been also made to clarify the superconducting 
pairing symmetry for (TMTSF)$_{2}$ClO$_{4}$. 
The superconducting transition temperature for (TMTSF)$_{2}($ClO$_{4}$)$_{1-x}$(ReO$_{4}$)$_{x}$ 
is suppressed by a tiny amount of non-magnetic impurities. \cite{Joo}  
Such a result may be evidence of the presence of nodes in the superconducting gap function.
Moreover, Takigawa {\it et al.} found that the nuclear magnetic relaxation rate lacks 
a coherence peak below $T_{c}$ together with 
a low temperature power-law behavior ($1/T_{1} \sim T^{3}$) \cite{Takigawa}.
These results suggest that the gap function has line-nodes at which the sign of the gap function changes. 
On the other hand, Belin and Behnia showed that 
the thermal conductivity rapidly decreases below $T_{\rm c}$.
The data leads to the absence of any nodal structures \cite{Belin}.


Recently, a technique rotating the applied magnetic field has been 
incorporated in thermal measurements such as heat capacity and thermal conductivity measurements 
to probe the gap structure 
including positions of gap nodes in more details \cite{Sakakibara,Matsuda}.
The angle-resolved heat capacity measurement is one of such advanced measurements, in which 
one detects details of the gap structure, especially locations of its nodes by measuring the oscillation 
of the heat capacity $C(\Vec{H})$ with respect to the 
applied magnetic field $\Vec{H}$ direction. 
The rotational dependence on the thermal conductivity 
is also a powerful tool to examine gap structures similarly.
In this paper, we propose Kramer-Pesch approximation (KPA) 
together with first-principles electronic-structure calculations as a new theoretical 
tool to analyze the advanced angle-dependent-measurement data \cite{NagaiPRL}.
KPA significantly exceeds 
the accuracy level of the previous Doppler shift approximation. 
We have calculated the density of states (DOS's) around a vortex by 
using anisotropic Fermi surfaces \cite{NagaiPRL,NagaiPRB,NagaiLow} obtained by 
first-principles calculations and 
compare calculated angle dependences of 
heat capacity with the experimental data.

 
Very recently, Yonezawa {\it et al.} reported that the oscillation curve of 
the heat capacity $C(\Vec{H})/T$ for (TMTSF)$_{2}$ClO$_{4}$ obtained by rotating the field 
becomes asymmetric with respect to the crystalline $a$-axis in the 
low temperature and low field range. \cite{Yonezawa}
In addition, kink structures are observed near this $a$-axis direction. 
They claimed,  based on the Doppler shift analysis,  that 
the asymmetries and the kink structures are clear evidence for the presence of line-nodes. 

On the other hand, a structural phase transition corresponding to an anion ordering 
was observed at 24 K for (TMTSF)$_{2}$ClO$_{4}$ \cite{Pouget, Leung, Pevelen}.
Shimahara proposed a nodeless fully-gapped $d$-wave superconductivity associated with the anion order \cite{Shimahara}.
Afterwards, several theoretical studies have suggested effective models with or without the anion ordering 
to discuss the pairing mechanism. \cite{Kuroki, Aizawa, Tanaka, Rozhkov, Belmechri,KurokiAoki}
However, there has been no first-principles calculation 
taking account of the anion ordering. 
Here, we emphasize that a trustful band calculation resolves effects of the anion ordering on the 
electronic structure.
%


The present paper has two objectives. The first one is 
to analyze the angle-resolved heat capacity for (TMTSF)$_{2}$ClO$_{4}$ with 
Fermi surfaces obtained by first-principles calculations. 
We identify the superconducting gap function by using an expression 
for the heat capacity on the basis of KPA in a low temperature 
and low magnetic field range. 
The KPA-based expression is applicable to 
various other unconventional superconductors. 
We calculate the heat capacity by assuming 
the $s$-wave, nodeless $d$-wave and nodal $d$-wave 
gap functions and compare the results with the measurement data.  
The second one is to examine the effects of the anion ordering on 
the electronic structure of (TMTSF)$_{2}$ClO$_{4}$.  
We clarify the electronic structure, especially Fermi surfaces 
through first-principles calculations using the measured structure parameters in 
the anion-ordered state. 



The rest of this paper is organized as follows. 
The quasiclassical approximation to describe 
the superconducting state is briefly introduced in Sec.~II. 
Then, the KPA calculation scheme based on the quasiclassical approach
is presented in Sec.~III. 
We derive a vortex solution using KPA and present an expression of the 
heat capacity around a vortex. 
The electronic structure of (TMTSF)$_{2}$ClO$_{4}$ by first-principles calculations is 
given in Sec.~IV, in which we display the band structure and Fermi surfaces. 
The calculation results on the angle-dependent heat capacity are shown in Sec.~V. 
The discussion and conclusion are, respectively, given in Sec.~VI. and VII.


\section{Quasiclassical theory of superconductivity}

In many BCS superconductors, the gap-amplitude is much smaller 
than the Fermi energy, $|\Delta| \ll E_{\rm F}$. 
In this case, one can properly use a 
quasiclassical approximation \cite{KopninText,Eilenberger68,Larkin68}. 
We consider the quasiclassical Green's function $\check{g}$ that has the matrix elements in the 
Nambu (particle-hole) space as 
\begin{align}
\check{g}(z,{\bm r},{\bm k}_{{\rm F}}) \equiv \left(\begin{array}{cc}
g & f \\
- \tilde{f} & -g\end{array}\right), 
\end{align}
which is a $2 \times 2$ matrix in the Nambu space and is a function of 
complex frequency $z$, Fermi wave-vector  ${\bm k}_{F}$, and point ${\bm r}$ in 
real space. 
We set $\hbar = k_{\rm B} = 1$ through this paper. 
The equation of motion for $\check{g}$ is written as 
\begin{equation}
- i \Vec{v}_{\rm F}(\Vec{k}_{\rm F}) \cdot \Vec{\nabla} \check{g} = \left[ 
z \check{\tau}_{3} - \check{\Delta}(\Vec{r},\Vec{k}_{\rm F}) ,\check{g}
\right], \label{eq:eilen}
\end{equation}
with Fermi velocity $\Vec{k}_{\rm F}$ and the commutator $[\check{a},\check{b}] = \check{a} \check{b} - \check{b} \check{a}$ 
supplemented by the normalization condition 
\begin{equation}
\check{g}^{2} = - \pi^{2} \check{1}. \label{eq:normal}
\end{equation}
Here, $\check{\Delta}$ is given by 
\begin{eqnarray}
\check{\Delta}(\Vec{r},\Vec{k}_{\rm F}) = 
\left(\begin{array}{cc}
0 & \Delta(\Vec{r},\Vec{k}_{\rm F}) \\
- \Delta^{\ast}(\Vec{r},\Vec{k}_{\rm F}) & 0\end{array}\right). 
\end{eqnarray}
We neglect the vector potential by confining ourselves in type II limit. 
Setting $z = \epsilon + i \eta$ with infinitesimal positive $\eta$, we obtain the 
retarded quasiclassical Green's function $\check{g}^{\rm R}$. 
In this paper, we use a special parameterization form of the quasiclassical Green's function 
to solve Eq.~
(\ref{eq:eilen}). \cite{Kato,Nagato,Higashitani, NagatoLow,SchopohlMaki,Schopohl} 
The solution $\check{g}$ of Eq.~(\ref{eq:eilen}) can be written as 
\begin{eqnarray}
\check{g} = \frac{- i \pi}{1 + a b} 
\left(\begin{array}{cc}1- ab & 2 i a \\
- 2 i b & -(1 - a b)\end{array}\right), 
\label{eq: riccati}
\end{eqnarray}
where $a$ and $b$ are the solutions of the following Riccati differential equations: 
\begin{align}
\Vec{v}_{\rm F}(\Vec{k}_{\rm F}) \cdot \Vec{\nabla} a &=  2 i  z  a 
- a^{2} \Delta^{\ast}  + \Delta , \label{eq:a} \\ 
\Vec{v}_{\rm F}(\Vec{k}_{\rm F}) \cdot \Vec{\nabla} b &= - 2 i z b 
 + b^{2} \Delta  - \Delta^{\ast} .\label{eq:b}
\end{align}
In the parameterization Eq.~(\ref{eq: riccati}), the normalization condition Eq.~(\ref{eq:normal}) 
is automatically satisfied. \cite{NagaiJPSJ} 


Since Eqs.~(\ref{eq:a}) and (\ref{eq:b}) contain $\Vec{\nabla}$ only through 
$\Vec{v}_{\rm F}(\Vec{k}_{\rm F}) \cdot \Vec{\nabla}$, they are reduced to a one-dimensional problem 
on a straight line, the direction of which is given by that of the Fermi velocity $\Vec{v}_{\rm F}(\Vec{k}_{\rm F})$.
We consider a single vortex along the $z_{\rm M}$ axis. 
Because of the translational symmetry along the $z_{\rm M}$ axis, the pair potential $\Delta$ does not depend on $z_{\rm M}$ 
in the Riccati equations (\ref{eq:a}) and (\ref{eq:b}), and hence 
the Riccati equations can be rewritten as 
 \begin{align}
v_{\rm F \perp}(\Vec{k}_{\rm F}) \frac{\partial}{\partial s} a &= 2 i z a 
- a^{2} \Delta^{\ast}(s,y,\Vec{k}_{\rm F})  + \Delta(s,y,\Vec{k}_{\rm F})  , \label{eq:as} \\ 
v_{\rm F \perp}(\Vec{k}_{\rm F}) \frac{\partial }{\partial s} b &= - 2 i  z b 
 + b^{2} \Delta(s,y,\Vec{k}_{\rm F})   - \Delta^{\ast}(s,y,\Vec{k}_{\rm F}) , \label{eq:bs}
\end{align}
where $v_{\rm F \perp}(\Vec{k}_{\rm F})$ is the amplitude of the vector $v_{\rm F \perp}(\Vec{k}_{\rm F})$ 
perpendicular to the $z_{\rm M}$ axis by projecting the 
Fermi velocity $\Vec{v}(\Vec{k}_{\rm F})$ and the coordinate $s$ ($y$) is along 
the direction parallel (perpendicular) to $\Vec{v}_{\rm F \perp}(\Vec{k}_{\rm F})$.  
For simplicity, we solve the Riccati differential equations under a given form of the pair function. 
The density of states is given by 
\begin{align}
N(\epsilon) &= \langle  \nu(\Vec{r} ,\epsilon ) \rangle_{\rm SP} , \\
\nu(\Vec{r},\epsilon) &= - \frac{1}{\pi}\int \frac{d S_{\rm F}}{2 \pi^{2} |\Vec{v}_{\rm F}|} {\rm Im}  \: \left( 
g^{\rm R}
\right). \label{eq:nu}
\end{align}
Here, $\langle \cdots \rangle_{\rm SP} \equiv \int_{0}^{r_{a}} r dr \int_{0}^{ 2 \pi}  d \alpha /(\pi r_{a}^{2})$ is 
the real-space average around a vortex 
where $r_{a}/\xi_{0} = \sqrt{H_{c2}/H}$ [$H_{c2} \equiv \Phi_{0}/(\pi \xi_{0}^{2}), \Phi_{0} = \pi r_{a^{2}} H$] and 
$dS_{\rm F}$ is the Fermi-surface area element.
By using $N(\epsilon)$, the low-temperature specific heat is given as 
\begin{align}
\frac{C(T)}{T} &= \int_{-\infty}^{\infty} \frac{d \epsilon}{T} 
\frac{\epsilon^{2}}{T^{2}}\frac{N(\epsilon) }{\cosh^{2} \left( \frac{\epsilon}{2 T}\right)}. \label{eq:ct}
\end{align}

\section{Kramer-Pesch Approximation}

We introduce Kramer-Pesch approximation (KPA) as an efficient method to analyze the angle-resolved experiments.  
We have shown that KPA gives the zero-energy density of states around a vortex consistent 
quantitatively with results of direct numerical calculations \cite{NagaiPRL}. 
In addition, the computational time required for KPA is almost the same as that for the Doppler Shift method, which 
is significantly less than that in direct numerical calculations. 
Furthermore, it is emphasized that KPA can calculate the density of states even in complicated Fermi surfaces 
without any heavy numerical computations. 
So far, we have actually examined various unconventional superconductors with 
the use of KPA \cite{NagaiPRL, NagaiJPSJLett,NagaiJPSJ, NagaiPRB,NagaiLow} .


In works based on KPA, there has been a different way 
in theoretical treatments on the vortex core.
Mel'nikov {\it et al.} presented an analytical solution describing the anomalous branches 
in a single vortex with arbitrary winding numbers by generalizing the Caroli-de Gennes-Matricon approach \cite{Melnikov}. 
They also demonstrated that the analytical solution on a single vortex is valid even in 
a higher energy range near the gap-amplitude.
Therefore, we incorporate the Mel'nikov's method to calculate the heat capacity. 
The Mel'nikov's method can be regarded 
as a perturbation with respect to both energy and imaginary part of 
the pair-function in the Riccati formalism.

Now, let us show the present scheme. 
First, we briefly mention the Doppler Shift method in the Riccati formalism for comparison. 
We separate the pair-potential $\Delta$ into the amplitude and the phase $\Phi(s,y,\Vec{k}_{\rm F})$ as 
\begin{align}
\Delta(s,y,\Vec{k}_{\rm F}) &= |\Delta(s,y,\Vec{k}_{\rm F}) | e^{i \Phi(s,y,\Vec{k}_{\rm F}) }.
\end{align}
Introducing $a = \exp( i \Phi) \tilde{a}$ and $b = \exp(- i \Phi) \tilde{b}$, 
the Riccati equation is written as 
\begin{align}
v_{\rm F \perp}(\Vec{k}_{\rm F}) \frac{\partial}{\partial s} \tilde{a} &= i \left( 2  z -
v_{\rm F \perp}(\Vec{k}_{\rm F}) \frac{\partial}{\partial s} \Phi
\right) \tilde{a} 
- \tilde{a}^{2} |\Delta|  + |\Delta|  , \label{eq:asD} \\ 
v_{\rm F \perp}(\Vec{k}_{\rm F}) \frac{\partial }{\partial s} \tilde{b} &= 
-i \left( 2  z -
v_{\rm F \perp}(\Vec{k}_{\rm F}) \frac{\partial}{\partial s} \Phi
\right)
\tilde{b} 
 + \tilde{b}^{2} |\Delta|   - |\Delta|. \label{eq:bsD}
\end{align}
Assuming $\partial \tilde{a}/\partial s = \partial \tilde{b} / \partial s = 0$, 
the equations can be exactly solved in an analytical way.
The solution is equivalent to that in the bulk region by replacing the energy $z$ 
with the Doppler shifted energy $z - (v_{\rm F \perp}/2) (\partial \Phi/ \partial s)$. 
The Doppler Shift method is an approximation neglecting the spatial variation of $|\bar{a}| = |a|$. 
Then, it breaks down near a vortex core \cite{Dahm,NagaiPRL}.

Next, we derive the vortex solution by using KPA. 
We write down a pair-potential around a vortex in the following form,  
\begin{align}
\Delta(s,y, \Vec{k}_{\rm F}) &= f(s,y) \Delta_{\infty} d(\Vec{k}_{\rm F}) e^{i \theta_{r}}, \\
&= f(s,y) \Delta_{\infty} d(\Vec{k}_{\rm F}) \frac{s+ i y}{\sqrt{s^{2} + y^{2}}} e^{i \theta_{v}(\Vec{k}_{\rm F})},
\end{align}
where $f(s,y)$ describes the spatial variation of the pair-potential.
Then, $f(0) = 0$, $\lim_{r \rightarrow \infty} f(r) = 1$, 
and $\Delta_{\infty}$ is a pair-potential in the bulk region.
$\theta_{r}$ denotes an angle around a vortex 
and $\theta_{v}$ does a direction of the projected 
Fermi velocity $\Vec{v}_{\rm F \perp}(\Vec{k}_{\rm F})$ \cite{NagaiJPSJ}. 
Introducing the variables written as 
\begin{align}
a & = \bar{a} e^{i \theta_{v}}, \\
b &=  \bar{b} e^{- i \theta_{v}}, \\
\Delta &= \bar{\Delta} e^{i \theta_{v}},
\end{align}
the Riccati equations are rewritten as 
 \begin{align}
v_{\rm F \perp}(\Vec{k}_{\rm F}) \frac{\partial}{\partial s} \bar{a} &= 2 i z \bar{a} 
- \bar{a}^{2} \bar{\Delta}^{\ast}+ \bar{\Delta}  , \label{eq:asb} \\ 
v_{\rm F \perp}(\Vec{k}_{\rm F}) \frac{\partial }{\partial s} \bar{b} &= - 2 i  z \bar{b} 
 + \bar{b}^{2} \bar{\Delta}  - \bar{\Delta}^{\ast} . \label{eq:bsb}
\end{align}
In KPA with the Riccati formalism as the previous paper 
\cite{NagaiPRL, NagaiJPSJLett,NagaiJPSJ, NagaiPRB,NagaiLow}, we expand 
$\bar{a}$ and $\bar{b}$ in Eqs.~(\ref{eq:asb}) and (\ref{eq:bsb}) with respect to 
the impact parameter $y$ and the complex frequency $z$. 
In this paper, we expand 
these variables with respect to the imaginary part of the pair-function $\bar{\Delta}$, 
in stead of $y$, and the complex frequency $z$ 
on the basis of the Mel'nikov's method. 
Then, $\Delta_{\rm R}$ and $\Delta_{\rm I}$ are defined as 
\begin{align}
\Delta_{\rm R} &= {\rm Re} \: \bar{\Delta} = f(s,y) \Delta_{\infty} d(\Vec{k}_{\rm F}) \frac{s}{\sqrt{s^{2} + y^{2}}}, \\
\Delta_{\rm I} &= {\rm Im} \: \bar{\Delta} = f(s,y) \Delta_{\infty} d(\Vec{k}_{\rm F}) \frac{y}{\sqrt{s^{2} + y^{2}}}. 
\end{align}
Following Refs.~\onlinecite{Kato} and \onlinecite{NagaiJPSJ}, 
we eventually obtain $\bar{a}$ and $\bar{b}$ as 
\begin{align}
\bar{a} &= \bar{a}_{0} + \bar{a}_{1} + {\cal O}(z^{2},\Delta_{\rm I}^{2}, z \Delta_{\rm I}), \\
\bar{b} &= \bar{b}_{0} + \bar{b}_{1} + {\cal O}(z^{2},\Delta_{\rm I}^{2}, z \Delta_{\rm I}), 
\end{align}
with 
\begin{align}
\bar{a}_{0} &= - {\rm sign} \: (d(\Vec{k}_{\rm F})) , \\
\bar{b}_{0} &=  {\rm sign} \: (d(\Vec{k}_{\rm F})), \\ 
\bar{a}_{1}(s) &= \frac{e^{u(s)}}{v_{\rm F \perp}(\Vec{k}_{\rm F})} \int_{- \infty}^{s} 
(2 i \bar{a}_{0} z + 2 i \Delta_{\rm I}(s'))e^{-u(s')} ds' \\
\bar{b}_{1}(s) &= \frac{e^{u(s)}}{v_{\rm F \perp}(\Vec{k}_{\rm F})} \int_{\infty}^{s} 
(-2 i \bar{b}_{0} z + 2 i \Delta_{\rm I}(s'))e^{-u(s')} ds'. 
\end{align}
Together with the help of the following function, 
\begin{align}
u(s) &= 2 \frac{|d(\Vec{k}_{\rm F})|}{v_{\rm F \perp}(\Vec{k}_{\rm F})}\int_{0}^{s}
\Delta_{\infty} f(s',y)\frac{s'}{\sqrt{s'^{2}+ y^{2}}} ds',
\end{align}
the quasiclassical Green's function is then written as 
\begin{align}
\check{g} &\sim \frac{- 2 \pi i}{\bar{a}_{1} \bar{b}_{0} + \bar{a}_{0} \bar{b}_{1}} 
\check{M}, \\
&= \frac{\pi v_{\rm F \perp}(\Vec{k}_{\rm F})}{C(y,\Vec{k}_{\rm F})}
\frac{e^{-u(s)}}{z -E(y,\Vec{k}_{\rm F})}\check{M}, \label{eq:g}
\end{align}
with 
\begin{align}
\check{M} &\equiv \left(\begin{array}{cc}
1 & i a_{0} \\
- i b_{0} & -1 \end{array}\right) , \\
C(y,\Vec{k}_{\rm F}) &\equiv \int_{-\infty}^{\infty} e^{-u(s')} ds', \label{eq:cc}\\
E(y,\Vec{k}_{\rm F}) &\equiv \frac{|d(\Vec{k}_{\rm F})| \Delta_{\infty} }{C(y,\Vec{k}_{\rm F})}
\int_{-\infty}^{\infty} f(s',y) \frac{y}{\sqrt{s'^{2} + y^{2}}}e^{-u(s')} ds'. \label{eq:ee}
\end{align}
The quasiclassical Green's function has the pole at $z = E(y,\Vec{k}_{\rm F})$, which 
is regarded as energy of quasi-particle. 
Substituting Eq.~(\ref{eq:g}) into Eqs. (\ref{eq:nu}) and (\ref{eq:ct}) and 
setting $z = \epsilon + i \eta$, 
the density of states is given as 
\begin{align}
N(\epsilon) &= \Bigl{\langle}  
\int \frac{d S_{\rm F}}{2 \pi^{2} |\Vec{v}_{\rm F}|} 
 \frac{v_{\rm F \perp}(\Vec{k}_{\rm F})}{C(y,\Vec{k}_{\rm F})}
e^{-u(s)} \delta(\epsilon -E(y,\Vec{k}_{\rm F}))
\Bigl{\rangle}_{\rm SP}. 
\end{align}
Thus, we obtain the heat capacity in the clean-limit ($\eta \rightarrow 0$) with 
the use of the KPA written as 
\begin{align}
\frac{C(T)}{T} &=
\Bigl{\langle}   \int \frac{d S_{\rm F} v_{\rm F \perp}(\Vec{k}_{\rm F}) }{2 \pi^{2} |\Vec{v}_{\rm F}| T^{3}} 
\frac{E(y,\Vec{k}_{\rm F})^{2} }{C(y,\Vec{k}_{\rm F})}
\frac{e^{-u(s)} }
{\cosh^{2} \left( \frac{E(y,\Vec{k}_{\rm F})}{2 T}\right)}  
\Bigl{\rangle}_{\rm SP}
. \label{eq:ctKPA}
\end{align}
On the other hand, we set the spatial variation of the pair-potential $f(s,y)$ as 
\begin{align}
f(s,y) &= \frac{r}{\sqrt{r^{2} +\xi_{0}^{2}}} = \frac{\sqrt{s^{2} + y^{2}}}{\sqrt{s^{2} + y^{2} + \xi_{0}^{2}}}.
\end{align}
With the use of this function, one can integrate Eqs.~(\ref{eq:cc}) and (\ref{eq:ee}): 
\begin{align}
C(y,\Vec{k}_{\rm F}) &= 2 \sqrt{y^{2} + \xi_{0}^{2}} K_{1}(r_{0}(y,\Vec{k}_{\rm F})) , \\
E(y,\Vec{k}_{\rm F}) &= |d(\Vec{k}_{\rm F})| \Delta_{\infty} \frac{y}{\sqrt{y^{2} + \xi_{0}^{2}}}
\frac{K_{0} (r_{0}(y,\Vec{k}_{\rm F}))}
{K_{1} (r_{0}(y,\Vec{k}_{\rm F}))},
\end{align}
with
\begin{align}
r_{0}(y,\Vec{k}_{\rm F}) &\equiv
\frac{2 |d(\Vec{k}_{\rm F})| \Delta_{\infty} }{v_{\rm F \perp}(\Vec{k}_{\rm F})} \sqrt{y^{2} + \xi_{0}^{2}}.
\end{align}
where, the function $K_{n}(x)$ is the modified Bessel function of the second kind.  


\section{Band structure}

Now, let us display the electronic structure.
In order to calculate the band structure, we employ 
a first-principles density-functional-calculation package VASP\cite{vasp}.
Among available options for the band calculations, we adopt GGA exchange-correlation energy \cite{pbe} 
and PAW method \cite{paw} due to their excellent computational 
performance as well as accuracy.
The lattice constants and atomic inner-coordinates refer to a measurement 
report, Ref.~\onlinecite{gallois}. 
The calculation self-consistent loops to obtain a converged electronic structure are 
repeated until the total energy difference becomes smaller than $10^{-6}$ eV. In the loops,  
$k$-points are taken as $10\times 5\times 5$, and the energy cut-off is set to be 500 eV.  
Once the electron density is obtained after the convergence, the energy bands 
are again calculated on finer $k$-points as 49$\times$23$\times$25
in order to determine the Fermi surfaces and the Fermi velocities accurately as much as possible.

\begin{figure}
\includegraphics[width = 6cm]{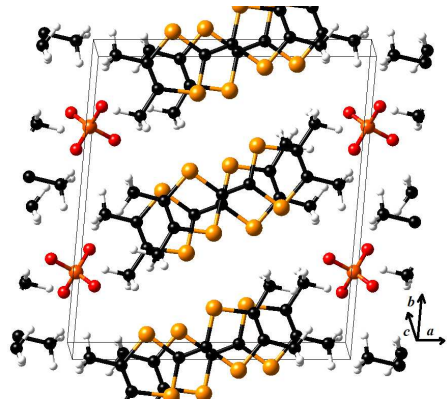}
\caption{\label{fig:structure}
(Color online) Crystal structure for (TMTSF)$_{2}$ClO$_{4}$.
}
\end{figure}

\begin{figure}
\includegraphics[width = 7cm]{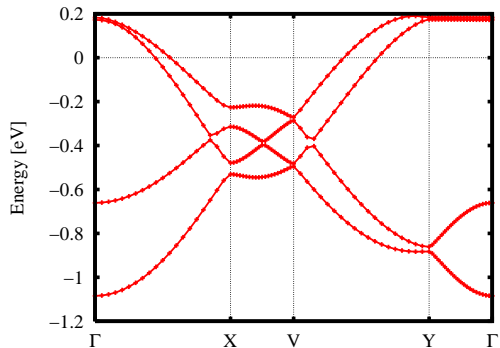}
\caption{\label{fig:band}
(Color online) Band structure for (TMTSF)$_{2}$ClO$_{4}$.
}
\end{figure}
\begin{figure}
  \begin{center}
    \begin{tabular}{p{60mm}p{20mm}}
      \resizebox{60mm}{!}{\includegraphics{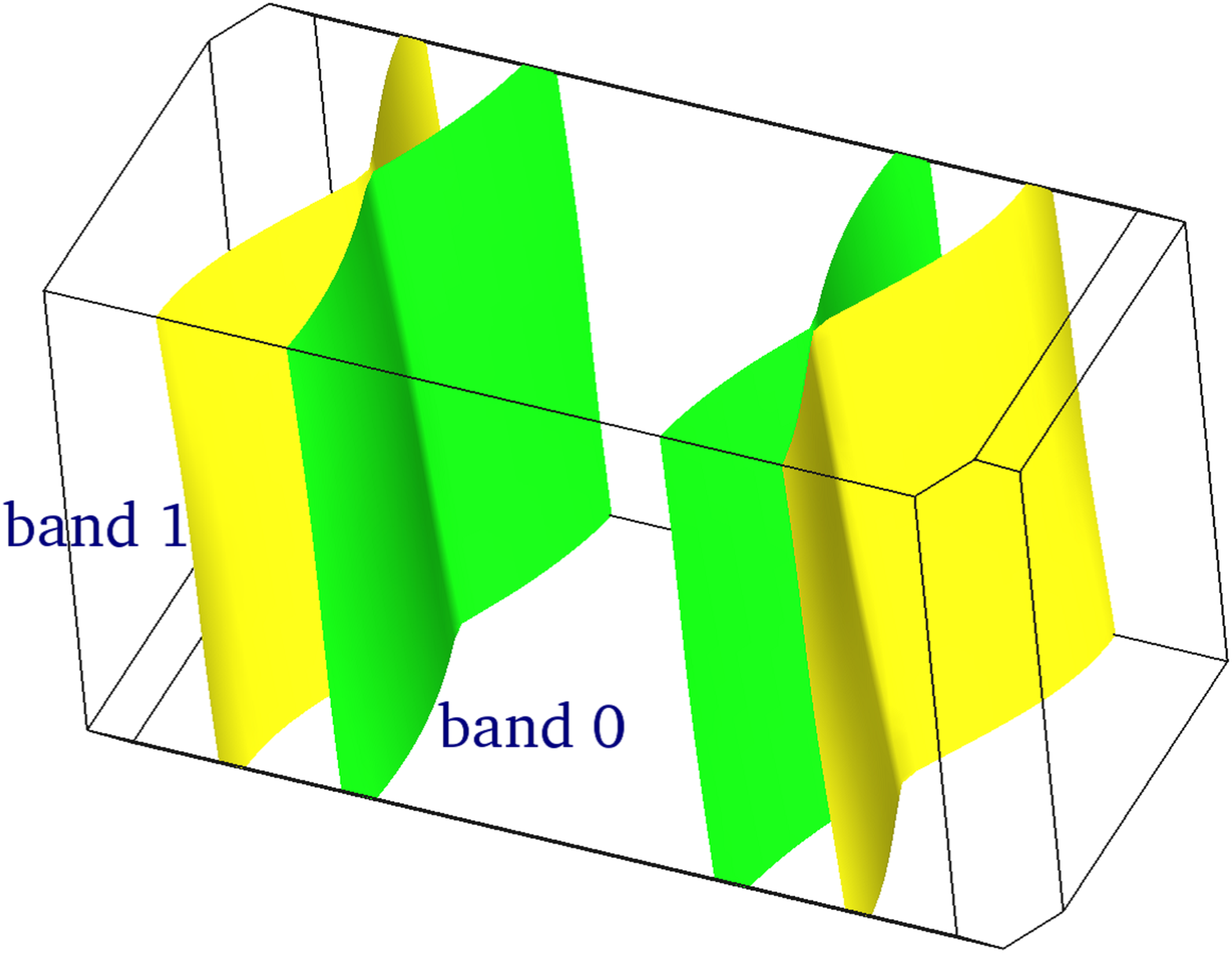}} &
      \resizebox{20mm}{!}{\includegraphics{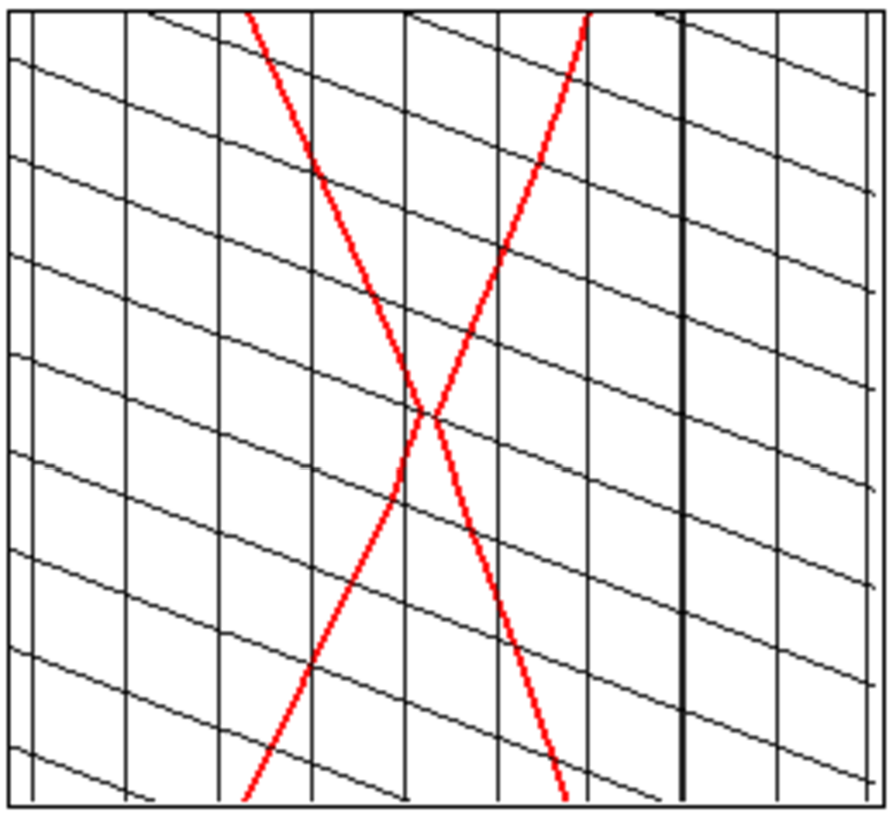}} 
    \end{tabular}
\caption{\label{fig:fermi}
(Color online) 
Fermi surfaces for (TMTSF)$_{2}$ClO$_{4}$. 
Inset: Closeup of the Fermi-surface crossing, sliced at $k_z=0$. The mesh
describes $k$-points actually used in the Fermi-surface calculation.
}
  \end{center}
\end{figure}

Since the employed structural parameters are measured 
at 7K, the data reflects the orientational ordering of the tetrahedral ClO$_{4}$ anions
in the crystal structure of (TMTSF)$_{4}$ClO$_{4}$. 
The ordered structure is displayed in Fig.~\ref{fig:structure}.
We obtain the band structure and Fermi surface for the structural parameters
as shown in Fig.~\ref{fig:band} and \ref{fig:fermi}. 
From Fig.~\ref{fig:fermi}, it is found that the two Fermi surfaces 
almost cross each other since the anion ordering gap is too small to resolve it 
in the standard scale. 
The tiny gap can be distinguished only by an enlarged 
scope as the inset of Fig.3. 
This result clearly suggests that the direction of the ClO$_{4}$ anion ordering 
does not have any significant effect on the Fermi surface structure contrary to 
the previous theoretical expectations. 
Thus, we would like to point out that any theoretical modelings originated from the anion ordering
are unlikely to consistently explain recent advanced experimental data.  


%

\section{Results}

We study the angle-resolved heat capacity, in which the applied magnetic field is rotated 
inside the basal $a$-$b$-plane. Three pairing symmetries are employed to test their 
matching with the angle-resolved experimental results.  
For simplicity, we assume that the vortex core is cylindrically isotropic and  
anisotropy of the critical magnetic field is not present, i.e., $H_{c2}(\phi) \sim H_{c2}$. 
On the analysis of the vortex core excitation, we set a spatial cutoff length $r_{a} = 5 \xi_{0}$, which is 
comparable to the neighboring vortex distance as the magnetic field $H \sim H_{c2}/25$. 
We take the $x$-axis ($y$-axis) parallel (perpendicular) to the $a$-axis. 
It is also noted that $y$-axis is parallel to the $b'$-axis introduced by the Ref.~\onlinecite{Wu} on $a$-$b$-plane and 
$z$-axis is perpendicular to the $a$-$b$-plane.

\subsection{$s$-wave gap function}

\begin{figure}
\includegraphics[width = 7cm]{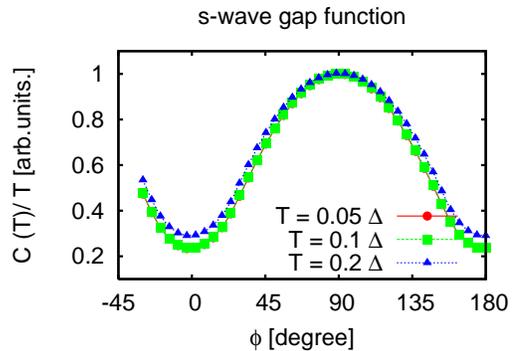}
\caption{\label{fig:sw}
(Color online) Angular dependence of the heat capacity rotating magnetic fields on $a$-$b'$ plane 
in the case of the $s$-wave gap function. 
}
\end{figure}

First, we examine a possibility of an isotropic $s$-wave gap function. 
In this case, the oscillation pattern of 
the angle resolved heat capacity suffers only the Fermi surface anisotropy. 
As shown in Fig.~\ref{fig:sw}, the heat capacity curve monotonically oscillates 
with the angle $\phi$ and shows the minima at $\phi =0$ reflecting the Fermi surface anisotropy.
The minima appear as the magnetic field direction is parallel to the $a$-axis.   
This oscillatory pattern is as one expects, but inconsistent with 
the latest measurement data in details.
In terms of the minima, we note that the Doppler shift method can 
not resolve even these minima, since 
the Doppler shift method can not describe the Fermi surface anisotropy in fully-gapped superconductors. 
Meanwhile, in the present scheme using KPA, the momentum $\Vec{k}_{\rm F}$ dependent 
kernel of the heat capacity (the integrand in Eq. (\ref{eq:ctKPA})) vanishes as the 
magnetic field is directed parallel to the Fermi velocity $v(\Vec{k}_{\rm F})$, since 
the projected Fermi velocity $v_{\rm F \perp}(\Vec{k}_{\rm F})$ then becomes zero. 
In the case of (TMTSF)$_{2}$ClO$_{4}$, the Fermi velocity is almost parallel 
to the $a$-axis on the whole Fermi surfaces because of the quasi-one-dimensionality. 


%

\subsection{Nodeless $d$-wave gap function} 

\begin{figure}
\includegraphics[width = 7cm]{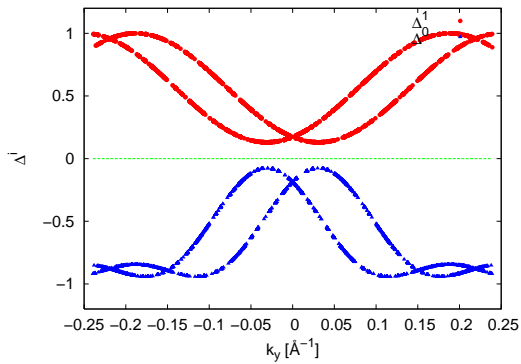}
\caption{\label{fig:gapfull}
(Color online) Gap functions on the Fermi surfaces in the case of the nodeless $d$-wave gap function.
}
\end{figure}

Next, we check a possibility of the nodeless $d$-wave gap function proposed by Shimahara. \cite{Shimahara}
We employ the nodeless $d$-wave gap function expressed as (see Fig.~\ref{fig:gapfull})
\begin{align}
\Delta^{0}(k_{x},k_{y},k_{z}) &= 
 a_{0} f^{0}(k_{x},k_{y} + b_{0}) 
, \\
 \Delta^{1}(k_{x},k_{y},k_{z}) &=
 -(a_{1} f^{1}(k_{x},k_{y}) + b_{1}),
 \end{align}
with
\begin{align}
 f^{0}(k_{x},k_{y}) &\equiv \cos \left( (k_{y}-k_{0})\frac{2 \pi }{g_{y}} \right)
 -\frac{1}{2} \cos \left( 2 (k_{y}-k_{0})\frac{2 \pi}{g_{y}} \right), \\
f^{1}(k_{x},k_{y}) &\equiv \cos \left((k_{y}-k_{0}) \frac{2 \pi}{g_{y}} \right), \\
 k_{0} &\equiv   - 0.031\: {\rm sign} \: (k_{x}). 
\end{align}
where, 
$a_{0,1}$ and $b_{0,1}$ denote normalization factors ($a_{0} = 10/26$, $b_{0} = -17/26$, $a_{1} = 10/23$ and 
$b_{1} = -3/23$), 
$g_{y}$ is the second element of the reciprocal lattice vector defined by $g_{y} = 2 \pi/(b \sin \gamma)$, 
$b$ denotes the crystal axis $b = 15.356$ {\AA} , $\gamma = 68.92^{\circ}$, and 
$k_{0}$ denotes the intersection of two Fermi surfaces. 
%
\begin{figure}
\includegraphics[width = 7cm]{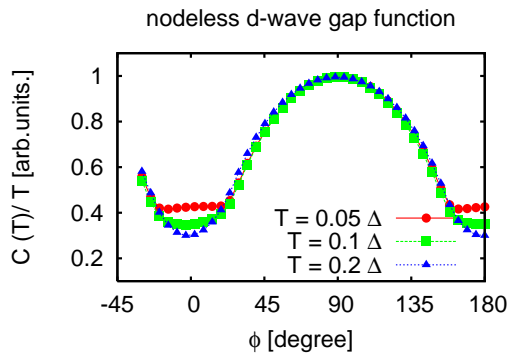}
\caption{\label{fig:nodeless}
(Color online) Angular dependence of the heat capacity rotating magnetic fields on $a$-$b'$ plane 
in the case of the nodeless $d$-wave gap function. 
}
\end{figure}
As shown in Fig.~\ref{fig:nodeless}, 
a slightly concave but almost flat-like curve showing the minimum at the 
center lies around the angles $-20^{\circ} < \phi < 20^{\circ}$. 
These symmetric kink-like curvature including the unclear minimum is 
due to anisotropy of the Fermi surface and the gap function. 
Close to $k_{y} = k_{0}$, the Fermi velocity continuously changes its direction around the $a$-axis, and 
the amplitude of the gap function gives the minimum with 
no change of the gap sign as shown in Fig.~\ref{fig:gapfull}. 
Therefore, the angle variation of the 
magnetic field around $a$-axis ($-20^{\circ} < \phi < 20^{\circ}$) 
is almost lost, i.e., DOS's of the quasiparticles with the small gap do not 
almost change with the angle. 
These results are inconsistent with the measurement data.

\subsection{Nodal $d$-wave gap function} 

\begin{figure}
\includegraphics[width = 7cm]{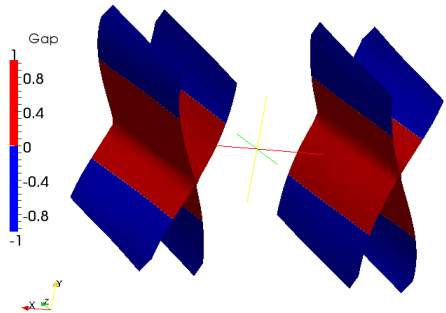}
\caption{\label{fig:gapnode}
(Color online) Schematic figure about the nodal line in the case of the nodal $d$-wave gap function I.
}
\end{figure}

Finally, we examine nodal $d$-wave gap functions.
The trial nodal $d$-wave gap functions are 
classified into three types expressed as 
\begin{align}
\Delta^{0}(k_{x},k_{y},k_{z}) &= 
\left\{ \begin{array}{ll}
f(k_{x},k_{y}) & ({\rm case \: I}) \\
f(k_{x},k_{y}) & ({\rm case \: II}) \\
1 & ({\rm case \: III}) \\
\end{array} \right. ,\\
\Delta^{1}(k_{x},k_{y},k_{z}) &=
\left\{ \begin{array}{ll}
f(k_{x},k_{y}) & ({\rm case \: I}) \\
1 & ({\rm case \: II}) \\
f(k_{x},k_{y}) & ({\rm case \: III}) \\
\end{array} \right. ,
\end{align}
with
\begin{align}
f(k_{x},k_{y}) &\equiv \cos \left(k_{y} \frac{2 \pi }{b_{y}} \right). 
\end{align}
The case I, whose nodes are on both the Fermi surfaces, 
is displayed in Fig.~\ref{fig:gapnode}, while the case II (III), whose nodes are only 
on the inner (outer) Fermi surface. 

%
%
\begin{figure}
\includegraphics[width = 7cm]{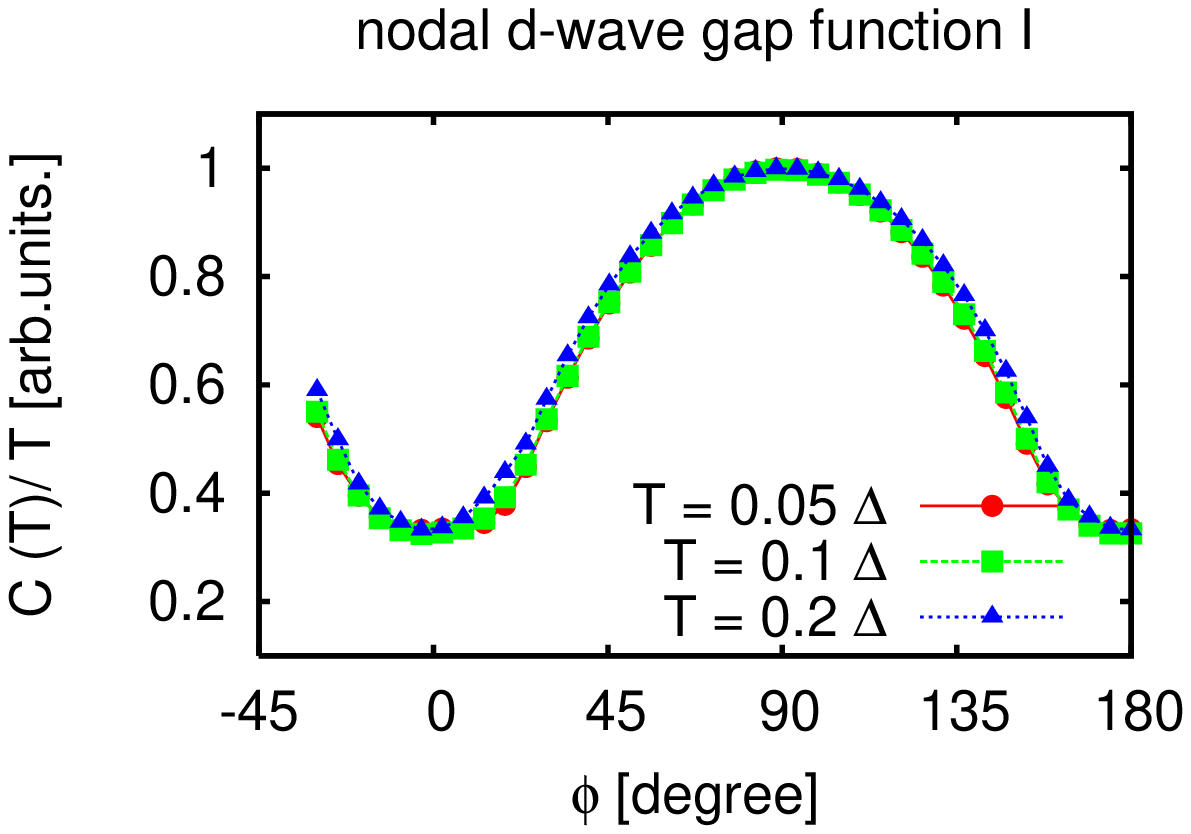}
\\
\includegraphics[width = 7cm]{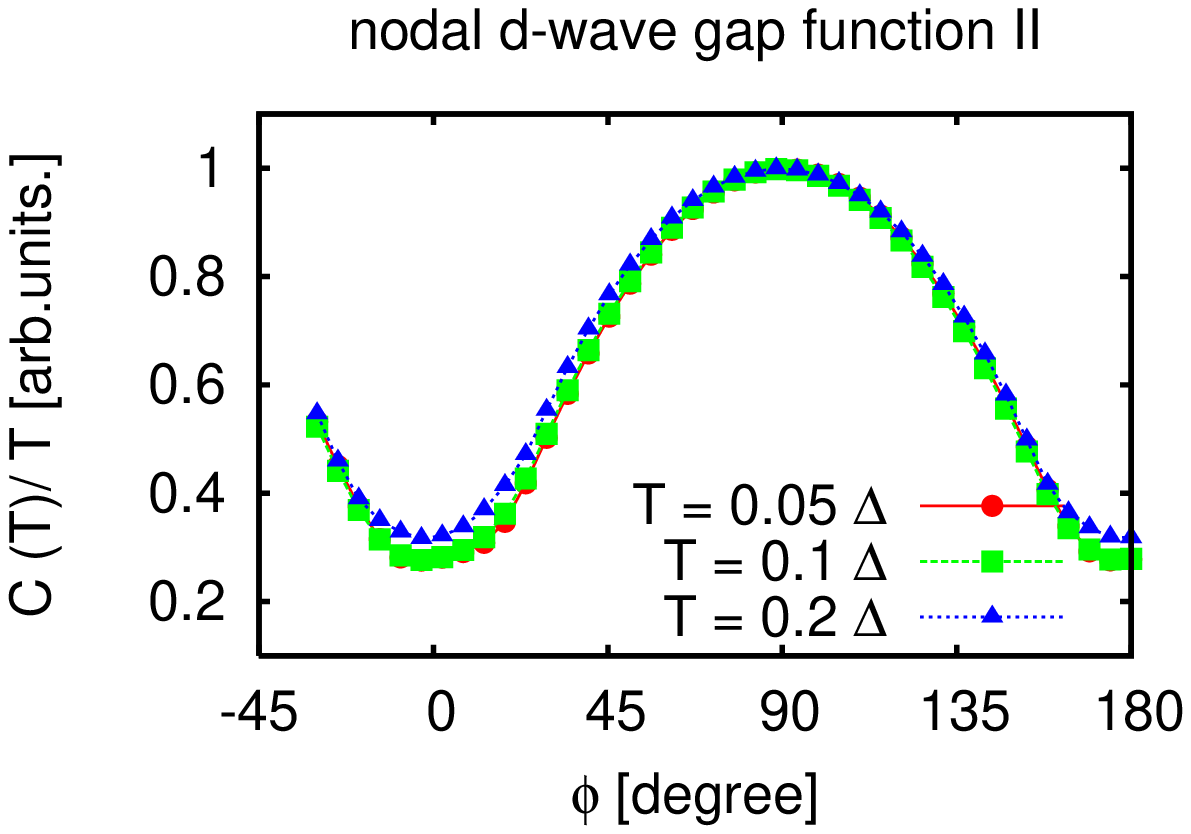}
\\
\includegraphics[width = 7cm]{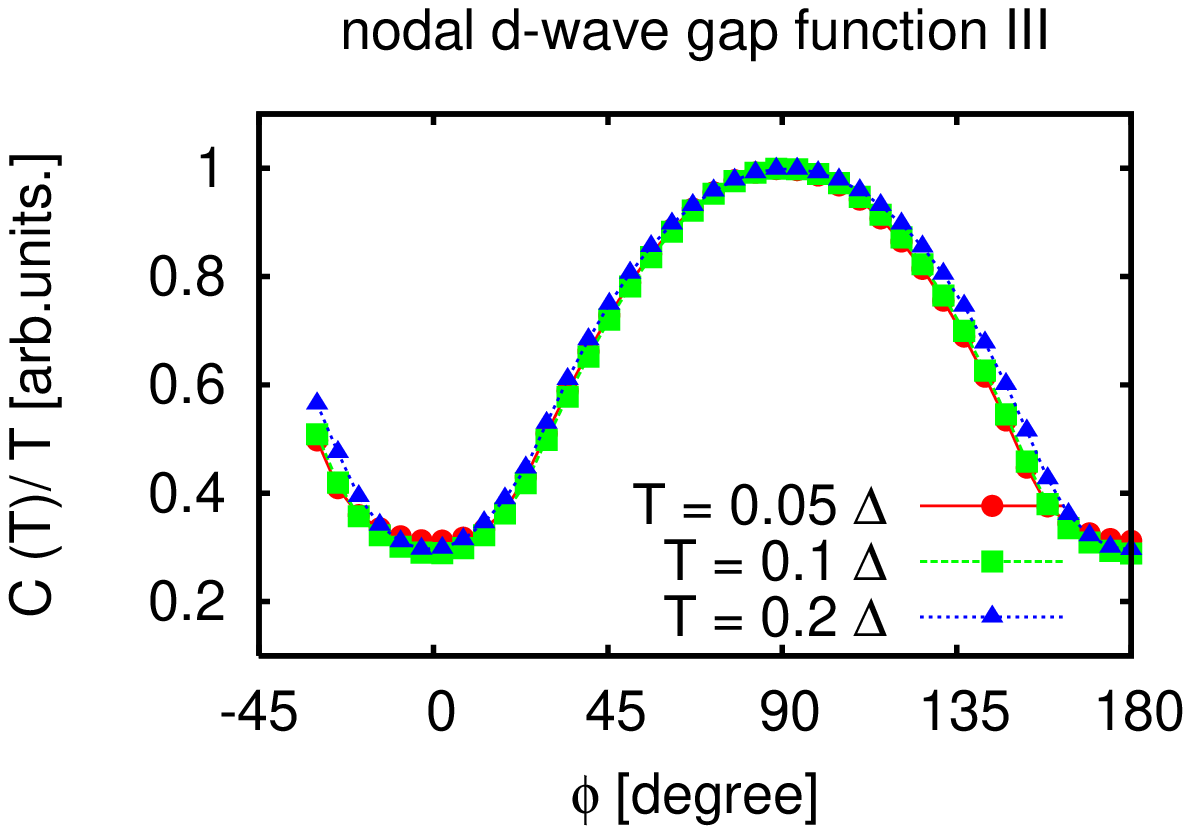}
\caption{\label{fig:nodal}
(Color online) Angular dependence of the heat capacity rotating magnetic fields on $a$-$b'$ plane 
in the case of the nodal $d$-wave gap function I, II, and III. 
}

\end{figure}
%
\begin{figure}
\includegraphics[width = 7cm]{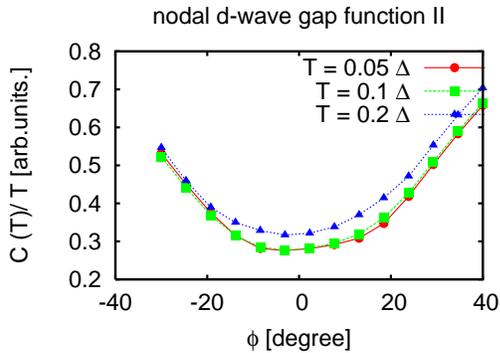}
\caption{\label{fig:nodalo0mag}
(Color online) Angular dependence of the heat capacity rotating magnetic fields on $a$-$b'$ plane 
in the case of the nodal $d$-wave gap function II. 
}
\end{figure}
%

As shown in Fig.~\ref{fig:nodal}, 
the asymmetric behavior with respect to the $a$-axis ($\phi = 0^{\circ}$) direction is found 
in the case of the nodal gap function II. 
Paying attention on the curve near $\phi = 0^{\circ}$ as shown in Fig.~\ref{fig:nodalo0mag}, 
the kink like structures are observed at around $\phi = \pm 15^{\circ}$ in an anti-symmetric manner.  
This calculation using the nodal gap function II is mostly
consistent with the experimental results.\cite{Yonezawa}
These results suggest that the superconducting gap nodes lie on the inner Fermi surface in the 0-th band.

\section{Discussion}

\begin{figure}
\includegraphics[width = 7cm]{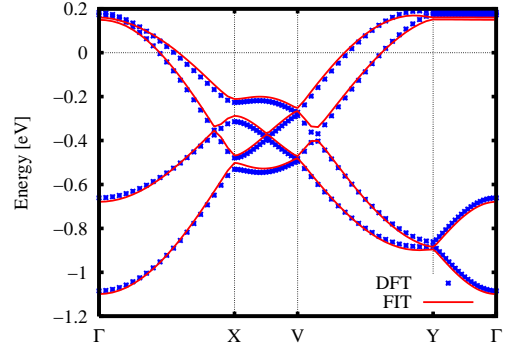}
\caption{\label{fig:tight}
(Color online) tight-binding fitted band structure (solid curves). the squares denote the first-principle 
band structure shown in Fig.~\ref{fig:band}.
}
\end{figure}

First, we mention the electronic structure including the Fermi surfaces.
In order to compare the present one with the previous tight-binding models \cite{Pevelen, Ishibashi}, we construct 
the tight-binding model based on our band calculation taking account of the anion ordering.
As shown in Fig.~\ref{fig:tight}, the obtained tight-binding model is equivalent with that by the 
band calculation. 
Following the Ref.~\onlinecite{Pevelen} in terms of the notation for the transfer integrals, 
the values we obtain are $t_{S_{1 {\rm A}}} = 280$ meV, $t_{S_{2 {\rm A}}} = 247$ meV, 
$t_{S_{1 {\rm B}}} = 269$ meV, $t_{S_{2 {\rm B}}} = 248$ meV, $t_{I_{1}} = -47.0$ meV, 
$t_{I_{2}} = -57.9$ meV, $t_{I_{3}} = 48.0$ meV, $t_{I_{4}} = -10.2$ meV, $t_{I_{5}} = 63.3$ meV, and 
$t_{I_{6}} = 3.98$ meV. 
These values are much more close to the values obtained by the band calculation 
without the anion ordering\cite{Ishibashi} than those obtained by phenomenologically 
taking account of the anion ordering. \cite{Pevelen} 
Therefore, the Fermi surfaces shown in Fig.~\ref{fig:fermi} are found to be qualitatively different from 
those obtained by the previous tight-binding model considering the anion ordering. 

Next, we discuss the assumption used in the gap examination, i.e., the excitation structure around vortex core is isotropic. 
In the Q1D superconductors, the superconducting gap amplitude may spatially vary around a vortex. 
However, low-lying quasiparticle excitations around a vortex are usually less affected by the variation of the gap amplitude 
than that of the superconducting phase.\cite{NagaiJPSJ,Hayashi}
Our expression on the heat capacity shown in Eq.~(\ref{eq:ctKPA}) can contain an anisotropic vortex core 
through $f(s,y)$.  
Therefore, it is a next-step checkpoint for the experimental consistency
 to investigate effects of the vortex core including the 
anisotropy. 

\begin{figure}
\includegraphics[width = 7cm]{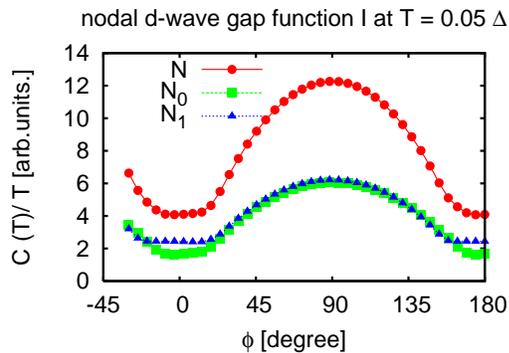}
\caption{\label{fig:t005}
(Color online) Angular dependence of the partial heat capacity rotating magnetic fields on $a$-$b'$ plane 
in the case of the nodal $d$-wave gap function I at $T = 0.05 \Delta$. $N_{i}$ is the partial heat capacity on the $i$-th band.
}
\end{figure}

Finally, we discuss the origin of the asymmetric behavior in the curves of the angle-resolved heat capacity. 
We calculate the partial heat capacity from each Fermi surface in the case of the nodal $d$-wave gap function I. 
As shown in Fig.~\ref{fig:t005}, 
the kink structures of the angle-dependent partial heat capacity are different on each Fermi surface 
at $T = 0.05 \Delta$. 
This difference originates from that in the Fermi velocity of the nodal quasiparticles on each Fermi surface. 
One finds that the asymmetric behaviors of the partial heat capacity are observed only in the case with the gap-nodes, since 
the distribution of the direction of the Fermi velocity does not have the strong asymmetry on the whole Fermi surfaces 
of (TMTSF)$_{2}$ClO$_{4}$. 
Therefore, the asymmetric kink structure of the curves measured in the angle-resolved heat capacity 
is a clear evidence that the gap function has nodes in 
(TMTSF)$_{2}$ClO$_{4}$. 

\section{Conclusion}
In order to examine the effects of the anion ordering and resolve the superconducting 
gap function in the organic superconductor (TMTSF)$_{2}$ClO$_{4}$, we performed first-principles calculations and developed the quasi-classical theory, respectively.
The first-principles calculation revealed that the anion ordering does not have any important role
on the Fermi surface shapes in contrast that the gap opening around the crossing point
was previously expected as a consequence of the ordering. The present calculation partly excludes the previous modeling based 
on the intuitive expectation.
On the other hand, using Kramer-Pesch approximation on the single 
vortex core excitation together with the Fermi surfaces obtained by the first-principles 
calculations, we constructed the formula calculating the angle-resolved heat capacity in 
the low-field range and compare the angle dependence obtained from various gap function models with the experimental results.
Consequently, we showed that the nodal $d$-wave gap function consistently explains the experimental results. 
Especially, it should be emphasized that only the nodal $d$-wave gap function can reproduce 
the axis asymmetry of the angle dependence.

\section*{Acknowledgment}
We thank N. Nakai, Y. Ota, and R. Igarashi for helpful discussions. 
We also thank S. Yonezawa for showing the latest experimental data.


\begin{thebibliography}{99}
\bibitem{Lee}
I. J. Lee, 
S. E. Brown, W. G. Clark, M. J. Strouse, M. J. Naughton, W. Kang, and P. M. Chaikin, 
Phys. Rev. Lett. {\bf 88}, 017004 (2002).   

\bibitem{Shinagawa}
J. Shinagawa, Y. Kurosaki, F. Zhang, C. Parker, S. E. Brown, D. J\'erome, J. B. Christensen, and K. Bechgaard, 
Phys. Rev. Lett. {\bf 98}, 147002 (2007). 



\bibitem{Kuroki}
K. Kuroki, J. Phys. Soc. Jpn., 
{\bf 75}, 051013 (2006). and references therein.



\bibitem{Aizawa}
H. Aizawa, K. Kuroki, Y. Tanaka, 
J. Phys. Soc. Jpn. {\bf 78}, 124711 (2009). 

\bibitem{Tanaka}
Y. Tanaka, and K. Kuroki, 
Phys. Rev. B {\bf 70}, 060502 (2004). 

\bibitem{Rozhkov}
A. V. Rozhkov, Phys. Rev. B {\bf 79}, 224501 (2009). 

\bibitem{Belmechri}
N. Belmechri, G. Abramovici, and M. H\'eritier, 
Europhys. Lett. {\bf 82}, 47009 (2008). 

\bibitem{Shimahara}
H. Shimahara, Phys. Rev. B {\bf 61}, R14936 (2000). 

\bibitem{KurokiAoki}
K. Kuroki, R. Arita, and H. Aoki, 
Phys. Rev. B {\bf 63}, 094509 (2001). 

\bibitem{Joo}
N. Joo, P. Auban-Senzier, C. R. Pasquier, D. J\'erome

\bibitem{Takigawa}
M. Takigawa, H. Yasuoka, G. Saito, 
J. Phys. Soc. Jpn. {\bf 56}, 873 (1987). 

\bibitem{Belin}
S. Belin and K. Behnia, Phys. Rev. Lett.  {\bf 79}, 2125 (1997). 

\bibitem{Sakakibara}
T. Sakakibara, A. Yamada, J. Custers, K. Yano, T. Tayama, H. Aoki, and K. Machida, 
J. Phys. Soc. Jpn. {\bf 76}, 051004 (2007). 

\bibitem{Matsuda}
Y. Matsuda, K. Izawa, I. Vekhter, 
J. Phys. Condens. Matter {\bf 18}, R705 (2006). 

\bibitem{NagaiPRL}
Y. Nagai and N. Hayashi, 
Phys. Rev. Lett. {\bf 101}, 097001 (2008).

\bibitem{NagaiPRB}
Y. Nagai, Y. Kato, N. Hayashi, 
K. Yamauchi and H. Harima, 
Phys. Rev. B {\bf 76}, 214514 (2007). 

\bibitem{NagaiLow}
Y. Nagai, N. Hayashi, Y. Kato, K. Yamauchi, and H. Harima, 
J. Phys. Conf. Ser. {\bf 150}, 052177 (2009). 

\bibitem{Yonezawa}
S. Yonezawa, Y. Maeno, and K. Bechgaard, 
International Conference on Science and Technology of Synthetic Metals 2010 (ICSM2010) 6Ax-09 (unpublished). 



\bibitem{Pouget}
J.-P. Pouget, G. Shirane, 
K. Bechgaard, 
J. M. Fabre, Phys. Rev. B {\bf 27}, 5203 (1985). 

\bibitem{Leung}
P. C. W. Leung, A. J. Schultz, H. H. Wang, 
T. J. Emge, 
G. A. Ball, D. D. Cox, and J. M. Williams, 
Phys. Rev. B {\bf 30}, 1615 (1984). 

\bibitem{Pevelen}
D. Le P\'evelen, J. Gaultier, Y. Barrans, 
D. Chasseau, F. Castet, and L. Duccase, 
Eur. Phys. J. B {\bf 19}, 363 (2001). 






\bibitem{KopninText}
N. Kopnin, {\it Theory of Nonequilibrium Superconductivity}  
(Clarendon, Oxford, 2001). 

\bibitem{Eilenberger68}G. Eilenberger, Z. Phys. {\bf 214}, 195 (1968).
\bibitem{Larkin68}A. Larkin and Yu. Ovchinnikov, Zh. Eksp. Teor. Fiz. {\bf 55}, 2262 (1968) [Sov. Phys. JETP {\bf 34}, 668 (1969)]. 

\bibitem{Kato}
Y. Kato, J. Phys. Soc. Jpn. {\bf 69}, 3378 (2000).

\bibitem{Nagato}
Y. Nagato, K. Nagai and J. Hara, 
J. Low Temp. Phys. {\bf 93}, 33 (1993).


\bibitem{Higashitani}
S. Higashitani and K. Nagai, 
J. Phys. Soc. Jpn. {\bf 64}, 
549 (1995). 

\bibitem{NagatoLow}
Y. Nagato, S. Higashitani, K. Yamada and K. Nagai, 
J. Low Temp. Phys. {\bf 103}, 1 (1996). 

\bibitem{SchopohlMaki}
N. Schopohl and K. Maki, 
Phys. Rev. B {\bf 52}, 490 (1995).

\bibitem{Schopohl}
N. Schopohl, arXiv:cond-mat/9804064 (unpublished). 



\bibitem{NagaiJPSJLett}
Y. Nagai, Y. Kato, and N. Hayashi, 
J. Phys. Soc. Jpn. {\bf 75}, 043706 (2006). 

\bibitem{NagaiJPSJ}
Y. Nagai, Y. Ueno, Y. Kato and N. Hayashi, 
J. Phys. Soc. Jpn. {\bf 75}, 104701 (2006). 



\bibitem{Melnikov}
A. S. Mel'nikov, D. A. Ryzhov, and M. A. Silaev, 
Phys. Rev. B {\bf 78}, 064513 (2008). 

\bibitem{Dahm}
T. Dahm, S. Graser, C. Iniotakis, and N. Schopohl, 
Phys. Rev. B {\bf 66}, 144515 (2002). 

\bibitem{vasp}
  G. Kresse and J. Hafner: Phys. Rev. B {\bf 47} (1993) RC558;
  G. Kresse and J. Furthm\"uller: Phys. Rev. B {\bf 54} (1996) 11169.
\bibitem{pbe}
J. P. Perdew, K. Burke, M. Ernzerhof: Phys. Rev. Lett. {\bf 77} (1996) 3865. 

\bibitem{paw}
P. E. Bl\"ochl: Phys. Rev. B {\bf 50} (1994) 17953;
G. Kresse and D. Joubert: Phys. Rev. B {\bf 59} (1999) 1758.

\bibitem{gallois}
B. Gallois, Ph.D thesis.

\bibitem{Wu}
W. Wu, I. J. Lee, and P. M. Chaikin, 
Phys. Rev. Lett. {\bf 91}, 056601 (2003). 

\bibitem{Ishibashi}
S. Ishibashi, A. A. Manuel, and M. Kohyama, 
J. Phys. Condens. Matter {\bf 11} 2279 (1999).

\bibitem{Hayashi}
N. Hayashi, M. Ichioka, and K. Machida, 
Phys. Rev. B {\bf 56}, 9052 (1997). 



\end{thebibliography}
\end{document}